\documentclass[11pt]{article}

\usepackage{times}
\usepackage[hyperindex=true,pageanchor=true,hyperfigures=true,backref=false]{hyperref} 
\usepackage[dvips]{color}
\usepackage{graphicx}
\usepackage{authblk}

\usepackage[utf8]{inputenc}

\usepackage{geometry}
\usepackage{setspace}

\usepackage{subfigure}

\usepackage{dsfont} 
\usepackage{amsmath}
\usepackage{amssymb}
\usepackage{amsthm}

\usepackage{listings}

\graphicspath{{figures/}}
\newcommand{\R}{\mathds{R}}
\newcommand{\mycdot}{\,\cdot\,}
\newcommand{\snorm}[1] {\Vert #1 \Vert}

\title{Forecasting with Historical Data or Process Knowledge under Misspecification: A Comparison}

\author[1]{Luke Bornn}
\affil[1]{Department of Statistics, University of British Columbia}
\author[2]{Marian Anghel}
\affil[2]{Information Sciences Group CCS-3, Los Alamos National Laboratory}
\author[3]{Ingo Steinwart}
\affil[3]{Department of Mathematics, University of Stuttgart}

\begin{document}
	
	\ifpdf
	\DeclareGraphicsExtensions{.pdf, .jpg, .tif, .png}
	\else
	\DeclareGraphicsExtensions{.eps, .jpg}
	\fi

\maketitle

\begin{abstract}
When faced with the task of forecasting a dynamic system, practitioners often have available historical data, knowledge of the system, or a combination of both.  While intuition dictates that perfect knowledge of the system should in theory yield perfect forecasting, often knowledge of the system is only partially known, known up to parameters, or known incorrectly.  In contrast, forecasting using previous data without any process knowledge might result in accurate prediction for simple systems, but will fail for highly nonlinear and chaotic systems.  In this paper, the authors demonstrate how even in chaotic systems, forecasting with historical data is preferable to using process knowledge if this knowledge exhibits certain forms of misspecification.  Through an extensive simulation study, a range of misspecification and forecasting scenarios are examined with the goal of gaining an improved understanding of the circumstances under which forecasting from historical data is to be preferred over using process knowledge.
\end{abstract}

\section{Introduction}

Whether it be predicting traffic flows next week, the longevity of nuclear stockpiles, or the climate next century, forecasting on all scales has become a crucial part of modern society.  How this forecasting of dynamic systems is performed, however, varies drastically.  A city engineer may use historical traffic volume data to predict upcoming flow; a nuclear scientist may use complex multi-physics based models to predict the lifespan of existing nuclear stockpiles; a climate scientist may use both complex physics-based models and historical climate data to predict long-range climate forecasts.  Typically, the choice of forecasting method is a function of the available resources, as well as the demands of the forecast itself.  Using the earlier examples, perhaps the city engineer is unaware of traffic-flow models, yet has considerable data on historic traffic flows; because of the United States' ban on nuclear testing, a nuclear engineer is faced with lack of data, and hence must rely on complex deterministic models to predict stockpile lifetime; in contrast, the climate scientist has available a wealth of physics-based models which describe our world, as well as data taken from historical records and inferred from ice cores and other methods.

What approach should be preferred?  If one has available perfect understanding of the underlying system (including noise mechanisms), then in theory perfect predictions are available.  Additionally, as one gains more and more data from a stationary system, they should be able to leverage this data to predict arbitrarily well (precluding the presence of noise).  However, it is less clear in general how these two choices break down as system-knowledge is misspecified or historical data becomes limited.  Highly nonlinear and chaotic (including meteorological) systems will exacerbate any incorrect knowledge, to the point of long-range forecasts becoming nearly useless.  It is therefore interesting to explore the effects of inaccurate information in the context of chaotic systems.

Knowledge-based approaches to forecasting are particularly prominent in fields connected to the natural sciences.  For instance, these models are used to forecast chemical reactions (\cite{mcquarrie1997physical}), determine strengths of engineering structures (\cite{sohn2004review}), and price options in financial markets (\cite{black1973pricing}).  Such models rely on a set of physical relationships expressed through mathematical formalism, usually differential equations.  These equations may be solved in closed form or through approximation methods; one such method is finite-element modeling, which reduces a system of partial differential equations to an approximate system of ordinary differential equations which may be solved with Euler's Method or related techniques.

In contrast to knowledge-based approaches, data-based approaches to forecasting rely solely on historical data.  For example, a retail outlet operator might predict next year's profits from previous years' records.  As another example, each year the Prognostics and Health Management Society holds a challenge contest, whereby participants are given a data file with only rough details of the data's origin.  From this, a model must be built which accurately predict a (withheld) testing set.  The participant who is best able to predict the testing set is awarded a cash prize and given a special invited session at the society's conference.  In situations where only data are available, one may select somewhat arbitrary parametric models (for instance linear regression or an autoregressive model) to reduce the dimensionality of the problem, or alternatively use non-parametric models such as support vector regression which make fewer assumptions on the distribution and relationships within the data.

While we have discussed only the previous two types of forecasting, it is worth noting that others exist.  For example, predictions of Oscar winners in a given year are a combination of data (the winners of the Golden Globes and other awards ceremonies) as well as expert knowledge of the films.  However, for the purpose of this exposition, we wish to compare knowledge and data-based forecasting methods in the forms described above.  Specifically, we will explore both of these two forms of forecasting under various levels of misspecification, testing when each method breaks down.

The organization of the paper is as follows: Section 2 discusses in detail prediction from historical data, focusing on Support Vector Machine (SVM) forecasters -- a non-parametric tool for predicting from data.  We move to knowledge-based approaches in Section 3, in particular focusing on filtering as an optimal tool for removing noise from dynamic systems.  The effects of misspecification are studied in Section 4, where an extensive simulation study is described.  Section 5 concludes the work with a description of lessons learned.

\section{Forecasting with Historical Data: SVM Forecasters}

Given a set of historical data, there is a potentially infinite range of possible models which may be used to forecast future data.  While the choice of model will be largely situation-dependent, such models may generally be classified as parametric or non-parametric.  The former relies on describing the data through a set of relationships determined by a set of parameters.  Such models include linear regression, generalized linear models, and autoregressive models.  In contrast, nonparametric models use historical data directly, comparing it with future data through some metric from which a model may be built.  This class includes neural networks and SVMs, the latter of which we now focus on in more detail.

Specifically, we begin by describing least squares support vector machines (LS-SVMs).
To this end, let $L:\R\times \R\to [0,\infty)$ be the least squares loss defined 
by $L(y,t):= (y-t)^2$, $y,t\in \R$. 
Moreover, let $X\subset \R^d$ be a non-empty subset and $k:X\times X\to \R$
be a {\em kernel\/}, i.e.~for every finite sequence $(x_1,x_2,\dots,x_n)\in X^n$, $n\geq 1$, 
the {\em kernel matrix\/} $K:= (k(x_i,x_j))_{i,j=1}^n$ is positive semi-definite. 
A particular choice, which we will consider throughout this work, is the 
{\em Gaussian RBF kernel} $k_\sigma:\R^d\times \R^d\to [0,\infty)$ defined by 
\begin{align*}
k_\sigma (x,x') := \exp\Bigl ( -\sigma^2 \snorm {x-x'}_2^2\Bigr)\, , \qquad \qquad x,x'\in \R^d, 
\end{align*}
where $\snorm \mycdot_2$ denotes the Euclidean norm and $\sigma>0$ is a free 
parameter called the {\em width}. It is well-known (\cite{aronszajn1950theory,steinwart2008support}),  that to every kernel on $X$ there is
a uniquely determined reproducing kernel Hilbert space (RKHS) $H$ on $X$, i.e., a Hilbert space consisting of functions that map
from $X$ to $\R$ such that $k(x,\mycdot)\in H$ and 
$$
f(x) = \bigl\langle f,k(x,\mycdot)\bigr\rangle
$$
for all $x\in X$ and $f\in H$. 

Let us now suppose that we have a finite sequence $D:=((x_1,y_1),\dots,(x_n,y_n))\in (X\times \R)^n$
of labeled samples. Then the LS-SVM finds the unique solution $f_{D,\lambda}\in H$ of 
the convex optimization problem
\begin{equation}\label{svm1}
  \arg\min_{f \in H}  \,\lambda \|f\|_H^2 +  \frac{1}{n} \sum_{i=1}^{n} \, L (y_i,f(x_i))\, ,
\end{equation}
where $\lambda>0$ is a regularization parameter and $H$ is a reproducing kernel Hilbert space  over $X$.
This approach has been widely considered in the literature, we refer to \cite{poggio1990theory,Wahba1990spline,Cucker2002mathematical} 
for the above formulation, and to \cite{Suykens2002least} for a formulation with an additional offset.
In particular, it is well known that the solution $f_{D,\lambda}\in H$ is given by 
\begin{align*}
f_{D,\lambda} = \sum_{i=1}^n \alpha_i k(x_i,\mycdot) ,
\end{align*}
where the vector $\alpha:= (\alpha_1,\dots,\alpha_n)\in \R^n$ is given by
\begin{equation}\label{alpha-direct}
\alpha = 2\lambda (K+\lambda I)^{-1} y\, ,
\end{equation}
where $K$ denotes the kernel matrix that corresponds to the samples  $(x_1,\dots,x_n)$, $I$ is $n$-dimensional diagonal matrix,
and $y:= (y_1,\dots,y_n)\in \R^n$ denotes the vector of labels. Unfortunately, computing $\alpha$ directly via (\ref{alpha-direct}) 
requires a matrix inversion,
which is an $O(n^3)$ operation and hence infeasible for larger datasets. On the other hand, \cite{keerthi2003algorithm}
proposed a optimization algorithm that finds $\alpha$ by 
maximizing the dual optimization problem of (\ref{svm1}) and that 
is usually computationally more efficient than the direct method.

\section{Forecasting with Process Knowledge: Filtering}

As discussed earlier, knowledge of the underlying process in the form of differential equations is often used to model complex systems.  In addition, one typically has a training set to initialize the model.  For instance, meteorological forecasts are continually tuned and adjusted given current measurements.  When such noisy measurements of the system are available, the removal of noise to find the true underlying state is known as data assimilation.  In the context of linear dynamic systems with Gaussian measurement noise, the standard choice is the Kalman filter.  For more general classes of dynamic systems, more general filtering techniques exist, as we describe now.

Whereas in the SVM forecaster framework data alone is used to model the system and hence provide predictions, filtering relies on knowledge of the process underlying the system as well as noisy measurements of this process.  Specifically, we assume that the system dynamics are known up to some parameter(s).  The underlying state-space model may be written as
\begin{align*}
	z_t|z_{t-1} &\sim p_{z,t}(z|z_{t-1})\\
	x_t|z_{t} &\sim p_{x,t}(x|z_t)
\end{align*}
where $z_t$ and $x_t$ denote the unobserved state and observation at time $t$, respectively; $p_{z,t}$ and $p_{x,t}$ are the state transition and measurement models, respectively.  Also, we assume a prior distribution $p(z_0)$ on $z_0$.  In the case of linearly additive noise, we may write this state-space model as
\begin{align}
	z_t &= f(z_{t-1}|\theta) + \eta_t \notag\\
	x_t &= h(z_t) + \epsilon_t.
	\label{eq:statespace}
\end{align}
Here both the stochastic noise $\eta_t$ and the measurement noise $\epsilon_t$ are mutually independent and identically distributed sequences with known density functions.  In addition, $f(z_{t-1}|\theta)$ and $h(z_t)$ are known functions up to some parameters $\theta$.

In order to build the framework on which to describe the filtering methodologies employed, we first frame the above state-space model as a recursive Bayesian estimation problem.  Specifically, we are interested in obtaining the posterior distribution
\begin{equation}
	p(z_{0:t}|x_{1:t})
	\label{eq:posterior}
\end{equation}	
where $z_{0:t} = \left\{ z_0, z_1, \dots, z_t \right\}$ and $x_{1:t} = \left\{ x_1,x_2,\dots,x_t \right\}$.  Often we don't require the entire posterior distribution, but merely one of its marginals.  For instance, we are often interested in the estimate of state given all observations up to that point; we call this distribution the filtering density and denote it as 
\begin{equation}
	p(z_{t}|x_{1:t}).
	\label{eq:filtering}
\end{equation}
By knowing this density, we are able to make estimates about the system's state, including measures of uncertainty such as confidence intervals.  Also, once an estimate of the state is obtained, the system may be propagated forward using $f$ to obtain forecasts.  If the functions $f$ and $h$ are linear and both $\eta_t$ and $\epsilon_t$ are Gaussian, Kalman filtering is able to obtain the filtering distribution in analytic form.  In fact it can be seen that all of the distributions of interest are Gaussian with means and covariances that can be simply calculated.  However, when the dynamics are non-linear or the noise non-Gaussian, the Kalman filter is only able to provide a linear approximation, hence we must turn to alternative methods.  

In the case of non-linear dynamics with Gaussian noise, the standard methodology is the extended Kalman filter, which may be considered as a nonlinear Kalman filter which linearizes around the current mean and covariance.  However, as a result of this linearization, the filter may diverge if the initial state estimate is wrong or the process is incorrectly modeled.  In addition, the calculation of the Jacobian in the extended Kalman filter can become tedious in high-dimension problems.  Because of this we turn to the unscented Kalman filter of \cite{wan2000unscented}, which approximates the nonlinearity by transforming a random variable instead of through a Taylor expansion, as the extended Kalman filter does.  The method uses a set of strategically chosen sample points which are evolved using the exact nonlinear dynamics, yet still capture the posterior mean and covariance accurate to the second order.  By employing a deterministic sampling technique known as the unscented transform (\cite{julier1997extension}), UKF selects a minimal set of sample points around the mean which are then propagated through the non-linear functions while recovering the covariance matrix.

When either the stochastic or measurement noise is non-Gaussian, Monte Carlo methods must be employed, in particular particle filters (\cite{doucet2001sequential}).  This Monte Carlo based filtering method relies on a large set of samples, called particles, which are evolved through the system dynamics with potentially non-Gaussian noise using importance sampling and bootstrap techniques.  At each time step the empirical distribution of these particles is used to approximate the distribution of interest and its associated features.  By sampling from some proposal distribution $q(z_{0:t}|x_{1:t})$ in order to approximate (\ref{eq:posterior}), we may use importance sampling with corresponding unnormalized weights
\begin{equation*}
	w_t = \frac{P(x_{1:t}|z_{0:t})P(z_{0:t})}{q(z_{0:t}|x_{1:t})}.
\end{equation*}
However, we typically wish to perform this estimation sequentially, and hence we can take advantage of the Markov nature of the state and measurement process along with proposal distributions of the form $q(z_{0:t}|x_{1:t}) = q(z_{0:t-1}|x_{1:t-1})q(z_{t}|z_{0:t-1},x_{1:t})$.  From this we obtain the recursive weight formula
\begin{equation*}
	w_t = w_{t-1}\frac{P(x_{t}|z_{t})P(z_{t}|z_{t-1})}{q(z_{t}|z_{0:t-1},x_{1:t})}.
\end{equation*}
This equation allows for the sequential updating of importance weights given an appropriate choice of proposal distribution $q(z_{t}|z_{0:t-1},x_{1:t})$, as well as simple calculation of the filtering density (\ref{eq:filtering}).  Since we can sample from this proposal distribution and evaluate the likelihood and transition probabilities, the particle filter simply involves generating a prior set of samples, evolving these samples forward with the proposal distribution, and subsequently calculating the importance weights.  In addition, to prevent particle degeneracy, we employ a resampling step to remove particles with low weight and multiply those with high weight (\cite{Douc2005comparison}).  

Often the problem of filtering isn't restricted to the estimation of state, but is also concerned with estimating some parameters $\theta$ of the dynamic model $f(z|\theta)$.  Further complicating matters, the only information we have about the state and the model parameters is the noisy measurements $\left\{x_t\right\}_{t\geq1}$.  While there are several approaches for solving this problem, we focus on dual estimation, namely the use of parallel filters to estimate state and model parameters (\cite{wan2000dual}).  Specifically, we use a state-space representation for both the state and parameter estimate problems.  While the state-space representation for the state is given in equation (\ref{eq:statespace}), the representation of the model parameters is given by
\begin{align*}
	\theta_t &= \theta_{t-1} + \nu_t \\
	x_t &= f(z_{t-1}|\theta_t) + \eta_t + \epsilon_t.
\end{align*}
Here $\eta_t$ and $\epsilon_t$ are as in (\ref{eq:statespace}), while $\nu_t$ is an additional iid noise term.  Thus one can run two parallel UKF filters for both state and parameters.  At each time step the current state estimate is used in the parameter filter and the current parameter estimate is used in the state filter.  Many more details on particle filtering may be found in the edited volume of \cite{doucet2001sequential}.

\section{Examining the Effects of Misspecification}

For the purpose of this article, we define misspecification broadly, including what may traditionally be termed ``underspecification.''  Firstly, mathematical models are often incapable of capturing the complete complexity of physical systems, and hence are often only an approximation.  Hence the difference between the mathematical model and underlying physical system might be considered misspecification, as might the use of incorrect parameters in these models.  Alternatively one might have to attempt to learn the parameters from available data.  For data-based models, misspecification can be understand in terms of lack of availability of data.  While a small, stationary system may be fully understood from a handful of data points, complex and chaotic systems might require considerable amounts of data just to gain a rough understanding of the relationships between variables. 

We now proceed to conduct an elaborate simulation study to gain further insight into the mechanisms by which misspecification causes these forecasting methods to break down.  Specifically, we look at data-based forecasting methods for range of amounts of historical data and training lengths, studying their prediction performance in both the immediate and distant future.  Similarly for knowledge-based forecasting methods, we explore misspecification in terms of unavailable training data, unknown parameters, and unknown noise structure.  We look at SVM forecasters and filtering, both discussed earlier and in the appendix, as our primary focus of study.  As will be seen, many of the lessons learned from these models may be generalized to their respective classes of forecasters.

To really stretch the limits of the forecasting methods, we test them on a chaotic system, specifically the Lorenz-3 system (\cite{lorenz1963deterministic}).  The 3-dimensional system, parameterized by the variables $\sigma, b$ and $r$ is described by the following differential equations
\begin{align*}
	\frac{dx}{dt} = \sigma(y-x),	\frac{dy}{dt} = x(r-z)-y,	\frac{dz}{dt} = xy-bz
\end{align*}
and for our purposes is truncated using Runge-Kutta.  From this underlying system, we add both system and observation noise (described in more detail later).

\subsection{Experimental Design}

We now describe the setup of the numerical experiments performed to compare the SVM forecaster and filtering approaches to forecasting.  All generated data is from the fore-mentioned Lorenz-3 system, with added noise as described later.  The two forecasters - SVM and filtering -- each have access to a noisy sequence of $T_p$ consecutive past observations, $X:=(x_{-T_p+1}, \ldots, x_{0}) $, called the training set. The goal of the forecaster is to predict the state $x_{T_f}$ of the dynamical system at future time $T_f$.  The prior information available to the forecasters is different.  The filter will have partial or complete knowledge about the dynamical system and the noise process, while the SVM forecaster will have no information about the dynamics or the observational noise process: instead, this forecaster uses additional past noisy observations (in addition to the training set $X$). Thus the SVM forecast is entirely a data-based model of the dynamics.

More specifically, the SVM forecaster has available to it a set of historical data used to ``learn'' the system.  We explore a wide range of lengths of historical data, ranging from $500$ to $16,000$ in two-fold increments.  For each of these lengths, we repeat the experiment several times and use the average performance.  Due to the computational intensity of training the SVM forecaster, we limit the number of repetitions for the larger historical data sets as shown in Table \ref{Ta:size_samples}.
\begin{table}
  \begin{center}
    \begin{tabular}{|l|r|}
      \hline
      Historical Data Size & Repetitions \\ \hline
      500 &		100 \\ \hline
      1000&		100 \\ \hline
      2000&		10 \\ \hline   
      4000&		10 \\ \hline     
      8000 &	        5       \\  \hline    
      16000&		5 \\ \hline       
    \end{tabular}
    \caption{For each historical data set size (first column) we repeat the
	forecasting experiment a number of times
	(second column); the forecasting performance is averaged for each size of the
	historical data set.}\label{Ta:size_samples}
  \end{center}
\end{table}
As for the filter, we explore 2 filters with different observation noise structures -- Gaussian and Laplace, to test the sensitivity of forecasting to the specified noise structure.  The Gaussian filter is fit using the UKF, while the Laplace filter is fit using particle filtering.  The general setup of the experiment is as follows:
\begin{enumerate}
\item Generate a long trajectory of the dynamical system.

\item Use a portion of the data to generate the historical data sets (see Table \ref{Ta:size_samples}) and build the SVM forecasters for each.

\item Generate $n=1000$ random indices $i$ that describe a sample $(X,x_{T_f})$ in the sense that $i$ points to the $x_{0}$ state. 

\item For each index $i$ representing a sample $(X,x_{T_f})$, compute the predictor $f(X)$ for each model and save it.  For each index $i$ we examine a range of past training set lengths $T_p = 5, 10, 20, 50, 100, 1000$ and future forecasting lengths $T_f = 1, 5, 10, 20, 30, 40, 50$.

\item For the SVM forecaster, the above is repeated for the different historical data sets (including replication) and the results averaged for each set.
\end{enumerate}

As discussed earlier, we explore a range of misspecification scenarios, starting from fully specified through to drastically misspecified.  Table \ref{Ta:dyn_sys} describes the 6 systems under consideration in detail.
\begin{table}
  \begin{center}
    \begin{tabular}{|c|c|c|c|c|c|c|}
	\hline
	 & & & & Stochastic & Observation & \\
	System & $\sigma$ & $b$ & $r$ & Noise & Noise & Filter Knowledge \\
	\hline
	DS1 & $10.0$  & $8.0/3.0$ & $28.0$  & None & $N(0,0.80)$ & All Known \\
	DS2 & $8.13$  & $0.53$    & $35.23$ & None & $N(0,1.13)$ & All Unknown \\
	DS3 & $12.18$ & $0.52$    & $13.14$ & None & $U([-0.5,0.5])$ & All Unknown \\
	DS4 & $4.82$  & $0.63$    & $20.09$ & None & $.5 N(0.1,0.25)$ & All Unknown \\
	    &         &           &         &      & $+ .5 N(-0.1,0.5)$ &  \\
	DS5 & $3.34$  & $0.54$    & $23.49$ & None & $.8 N(0.1,0.25)$ & All Unknown \\
	    &         &           &         &      & $+ .2 U([-0.1,0.5])$ &  \\
	DS6 & $9.57$  & $3.04$    & $27.32$ & $N(0,0.1)$ & $0.5 \{\exp(1)/4\}$ & All Unknown \\
	    &         &           &         &            & $+ .5 * \{-\exp(1)/4\}$ & \\
	\hline      
    \end{tabular}
    \caption{Description of the dynamic systems used in the
forecasting experiments. Except for $DS1$, where both the Lorenz
parameters and the observational noise parameters are known, the system and noise parameters are not known to the
forecaster. Here, $N(m,\sigma)$ denotes a Gaussian distribution with mean
$m$ and standard deviation $\sigma$, $U([a,b])$ is uniform noise in
the range $[a,b]$, $\exp(\lambda)$ denotes the exponential distribution of
parameter $\lambda$, while $p_1 N(m_1,\sigma_1) + p_2 N(m_2,\sigma_2)$
denotes a mixture of Gaussian distributions where $p_1$ is the probability
of sampling from $ N(m_1,\sigma_1) $ and $p_2$ is the probability of
sampling from $N(m_2,\sigma_2)$; a similar interpretation holds for
the other mixture distributions in the table. }\label{Ta:dyn_sys}
  \end{center}
\end{table}
We see from this table that DS1 is quite favorable to the filter, as all aspects of the system are known.  DS2 through DS6 contain various levels of noise, and the filter attempts to learn the  system in each case.  Worthy of note is DS6, which in addition to observation noise contains stochastic noise.  Details of the tuning of each forecasting method are deferred to the appendix.

\subsection{Results of Experiment}

We now proceed to discuss the results of the experiment in order of the dynamic systems studied.  For each system we plot the $1$, $10$, $30$, and $50$ step ahead forecasting RMSE.  Due to space constraints and the similarities between DS3, DS4, and DS5, we exclude the plots of DS4 and DS5.  Because the maximum embedding length of the SVM forecaster is set to $20$, we indicate performance for larger training sizes with a dashed line, as all training data in addition to the first $20$ is not used by the forecaster.

\begin{figure}
  \centering
      \includegraphics[width=0.95\textwidth]{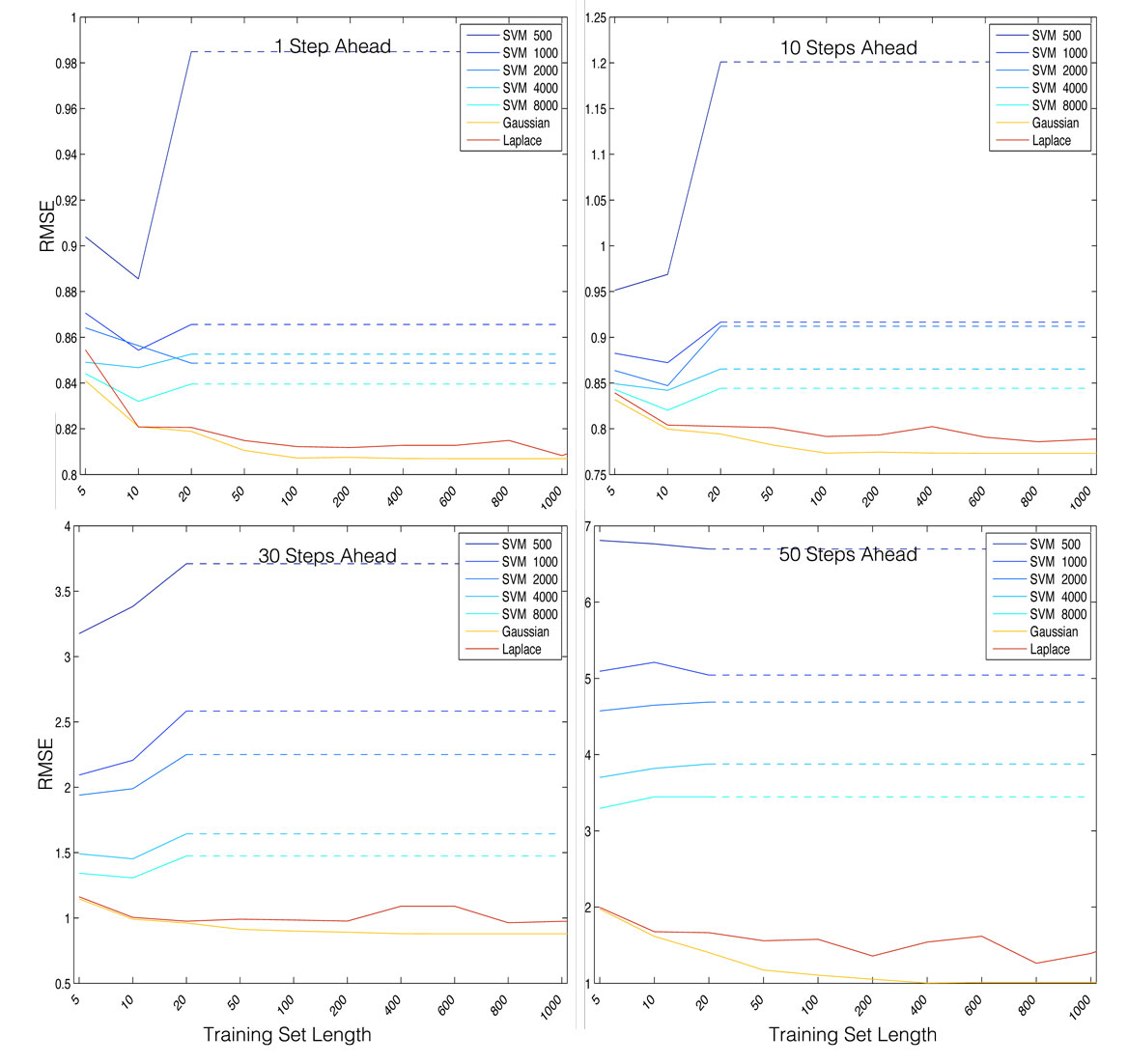}
  \caption{Root mean squared error of forecasting methods from DS1 vs. length of training set.  Four panes represent forecasting $1$, $10$, $30$, and $50$ steps ahead.}\label{fig:DS1}
\end{figure}

We first look at DS1 (Figure \ref{fig:DS1}), in which the filter has complete knowledge of the system, and hence the filter can be built  with these quantities fixed.  Since the noise and parameters are fixed in the filter, all that is needed is time to locate the state.  As such, we see slight improvements in prediction performance with increases in embedding length, as the state is able to be located more accurately and precisely.  As expected, we find the RMSE of the filtering method to dominate the SVM forecaster, with the Gaussian filter (which uses the correctly specified noise) outperforming the Laplace filter, which has incorrectly specified noise structure.  The improvement of using knowledge via the filter in this case is more pronounced for larger forecasting lengths, as the SVM forecaster is not able to accurately capture dynamics of this range.  Interestingly, the SVM forecaster seems to do no better with additional training data ($5$, $10$, or $20$).  Where significant gains are made by this forecaster is with larger historical data sets.

\begin{figure}
  \centering
      \includegraphics[width=0.95\textwidth]{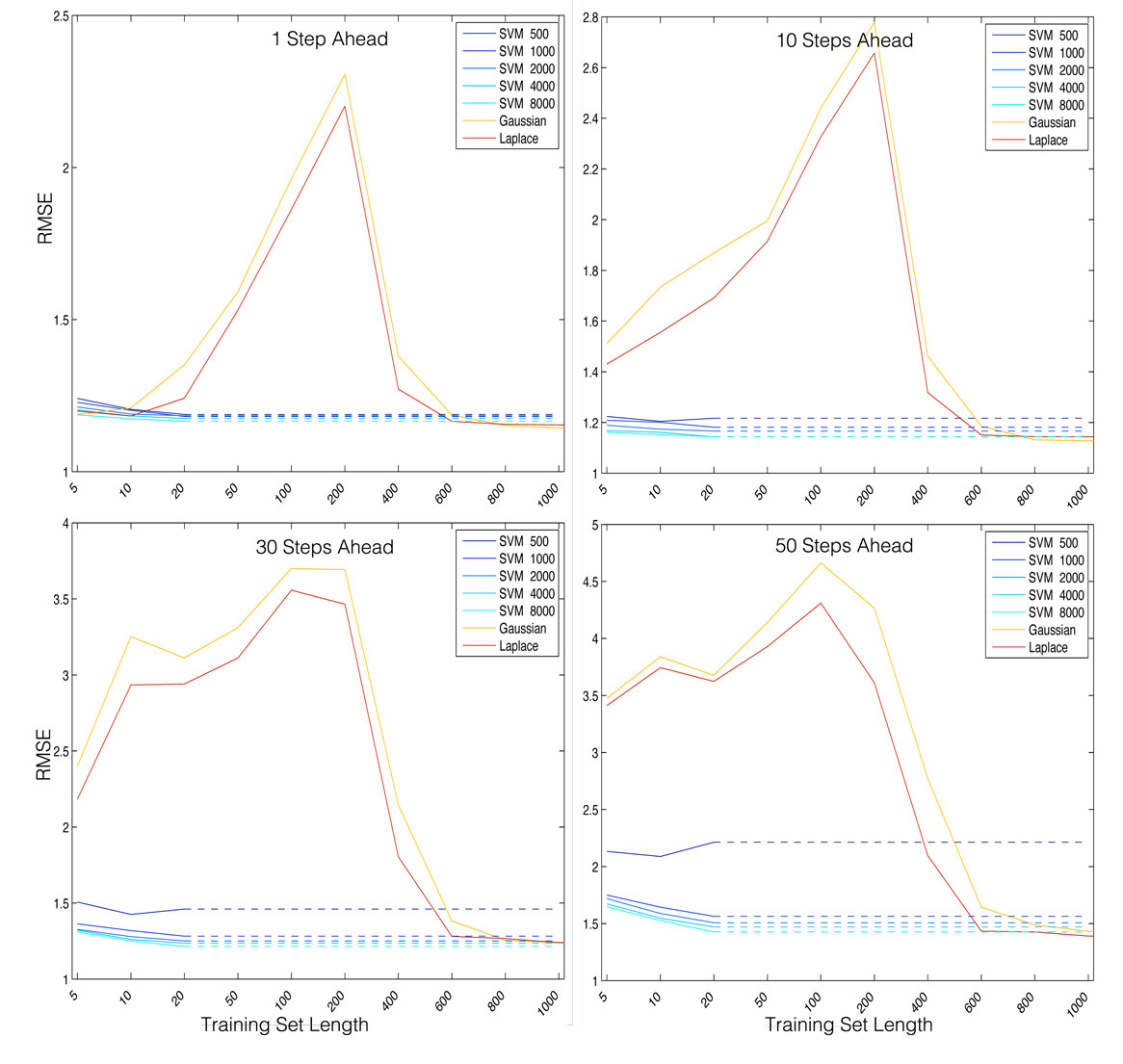}
  \caption{Root mean squared error of forecasting methods from DS2 vs. length of training set.  Four panes represent forecasting $1$, $10$, $30$, and $50$ steps ahead.}\label{fig:DS2}
\end{figure}

\begin{figure}
  \centering
      \includegraphics[width=0.95\textwidth]{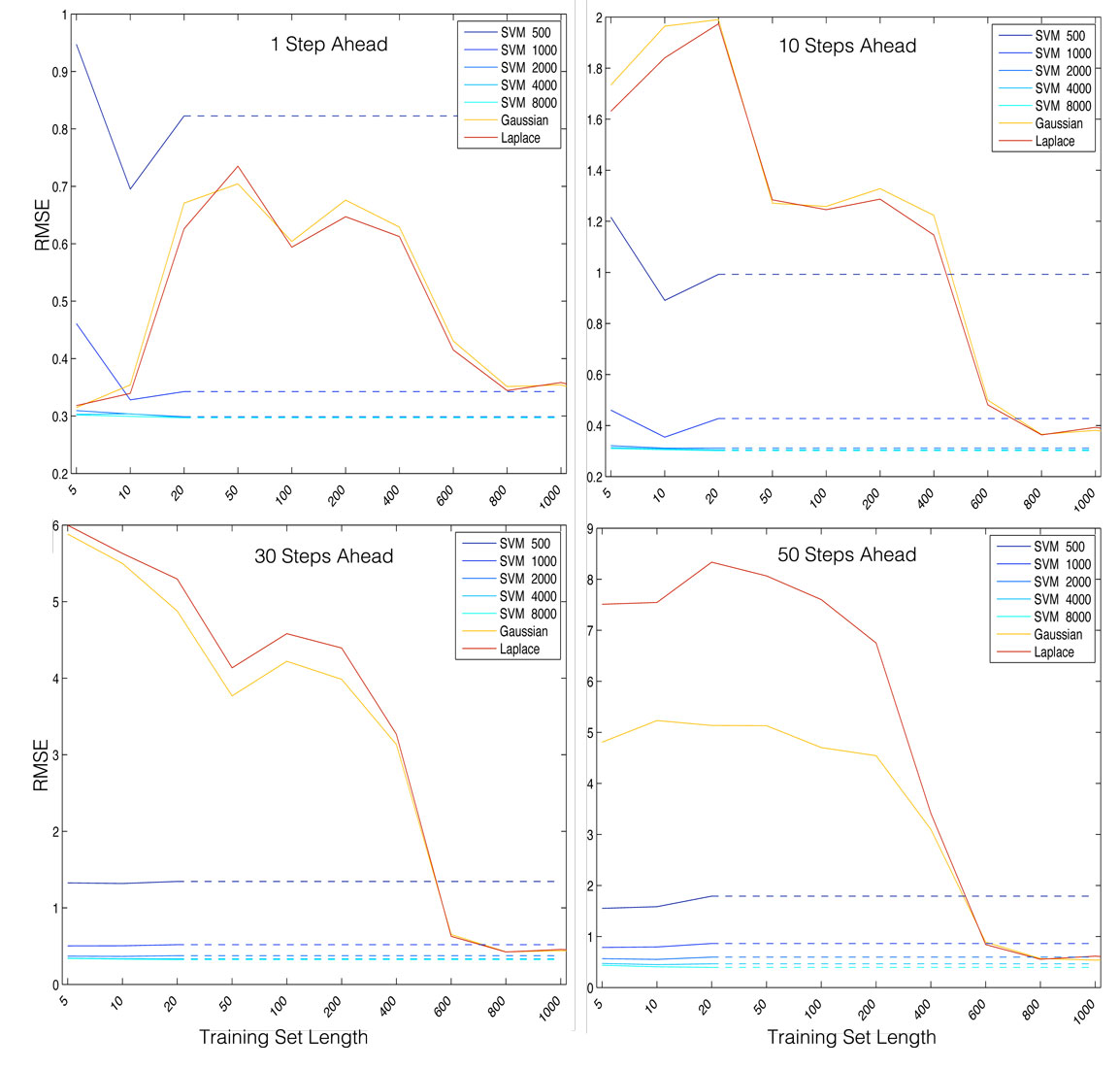}
  \caption{Root mean squared error of forecasting methods from DS3 vs. length of training set.  Four panes represent forecasting $1$, $10$, $30$, and $50$ steps ahead.}\label{fig:DS3}
\end{figure}

Looking at DS2 (Figure \ref{fig:DS2}), we see some interesting features of the forecasting performance.  Specifically, because we initialize the state at the first observation, the filter does not have much time to diverge with small training sets, and hence for small training sets and short forecasting lengths, performance is excellent.  However, as more training data is observed and the model attempts to learn the system parameters, the estimate of state is compromised, resulting in worsened forecasts.  However, as we see $600+$ training data, both state and parameters are found, and hence prediction capabilities equal or surpass the SVM forecaster.  Additionally, note that the Laplace filter (fit using particle filtering) outperforms the Gaussian.  This is likely due to the larger system variance, for which the Laplace noise is able to sufficiently accommodate, whereas the specified standard deviation of the Gaussian filter is smaller than the actual observation noise ($0.8$ vs. $1.13$) and hence outlying points are too influential.  Similar results are observed for DS3 through DS5 (DS3 shown in Figure \ref{fig:DS3}), however the additional noise misspecification results in the SVM forecasters with long historical data sets to uniformly dominate the filters.  

\begin{figure}
  \centering
      \includegraphics[width=0.95\textwidth]{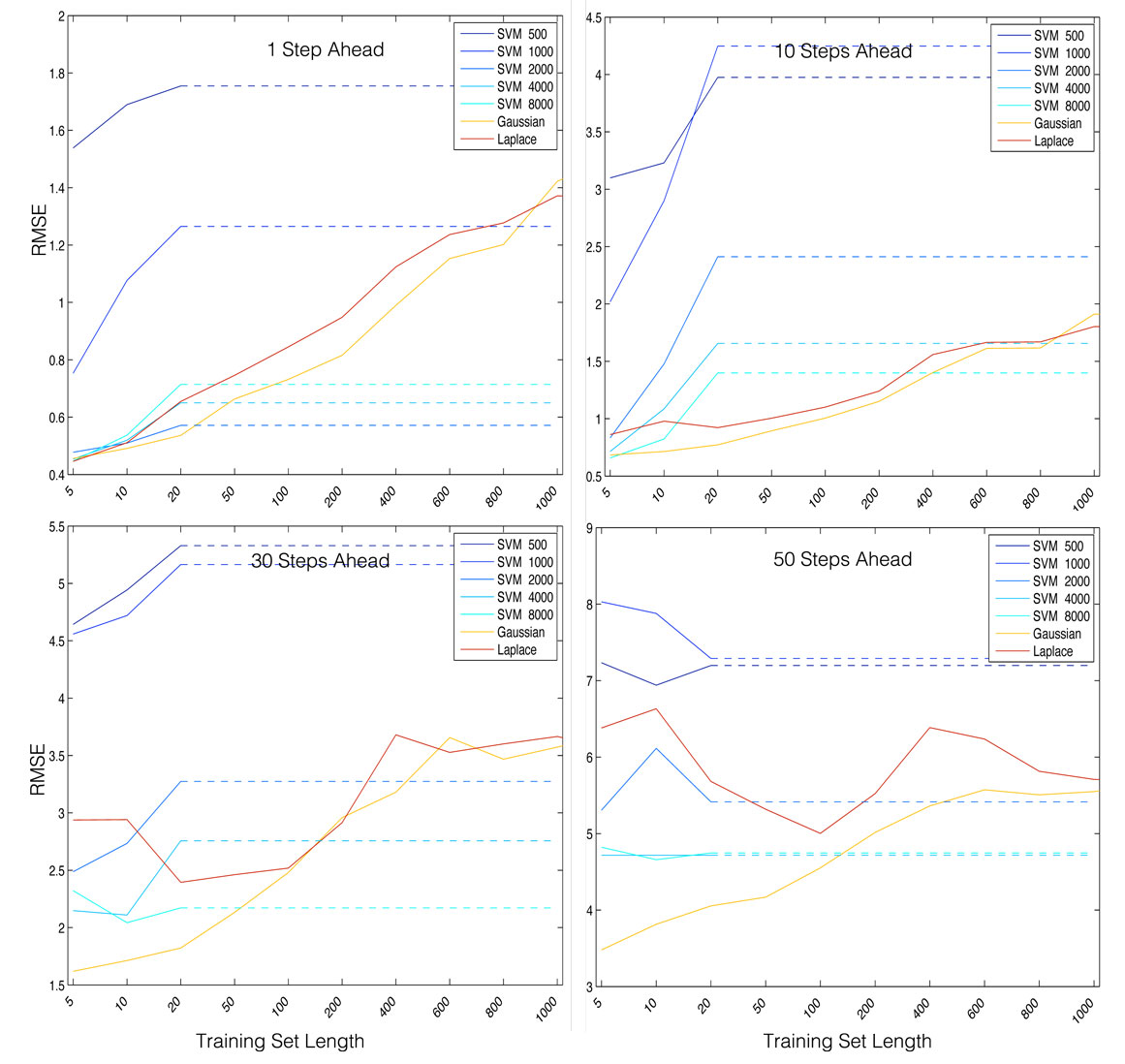}
  \caption{Root mean squared error of forecasting methods from DS6 vs. length of training set.  Four panes represent forecasting $1$, $10$, $30$, and $50$ steps ahead.}\label{fig:DS6}
\end{figure}

As we add in stochastic noise in DS6 (Figure \ref{fig:DS6}), we observe very interesting features.  Firstly, the SVM forecaster becomes worse with increased training set length for short-range forecasts.  Because of the relative smoothness of the system, this indicates that even with stochastic noise the system has little chance to deviate significantly.  While it is initially quite curious that the filters outperform the SVM forecasters for small training set, this is in fact due to the initialization of the filters.  Specifically, the prior on the parameter is set at the values for DS1, specifically $10.0$, $8.0/3.0$, and $28.0$, whereas the DS6 system parameters are $9.57$, $3.04$, and $27.32$ which is remarkably close.  Thus for short forecasting lengths, the prior is quite accurate, and hence forecasts are quite good.  However, as more training data is observed, the filters diverge from the true parameters and state, and hence we see forecasting degrades to roughly the average of the SVM forecasters.  Thus in this example, we see that barring lucky initialization, SVM forecasters are preferable if sufficient historical data is available.  However, because of the limited information in the historic data due to the presence of stochastic noise, the filter may outperform the SVM for moderate historical data.

\begin{figure}
     \centering
     \subfigure[Model parameter convergence for different initial conditions]{
          \label{fig:paramconv}
          \includegraphics[height=80mm, width=80mm]{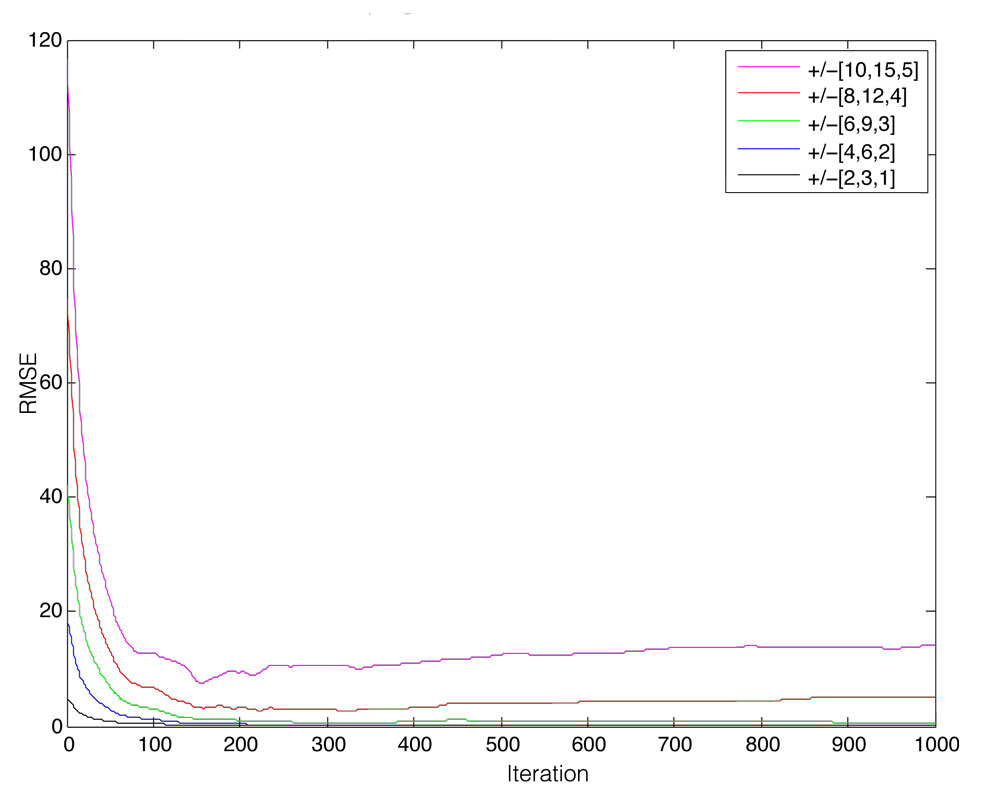}}
     \subfigure[Example of model parameter convergence and divergence]{
          \label{fig:paramcomp}
          \includegraphics[height=80mm, width=80mm]{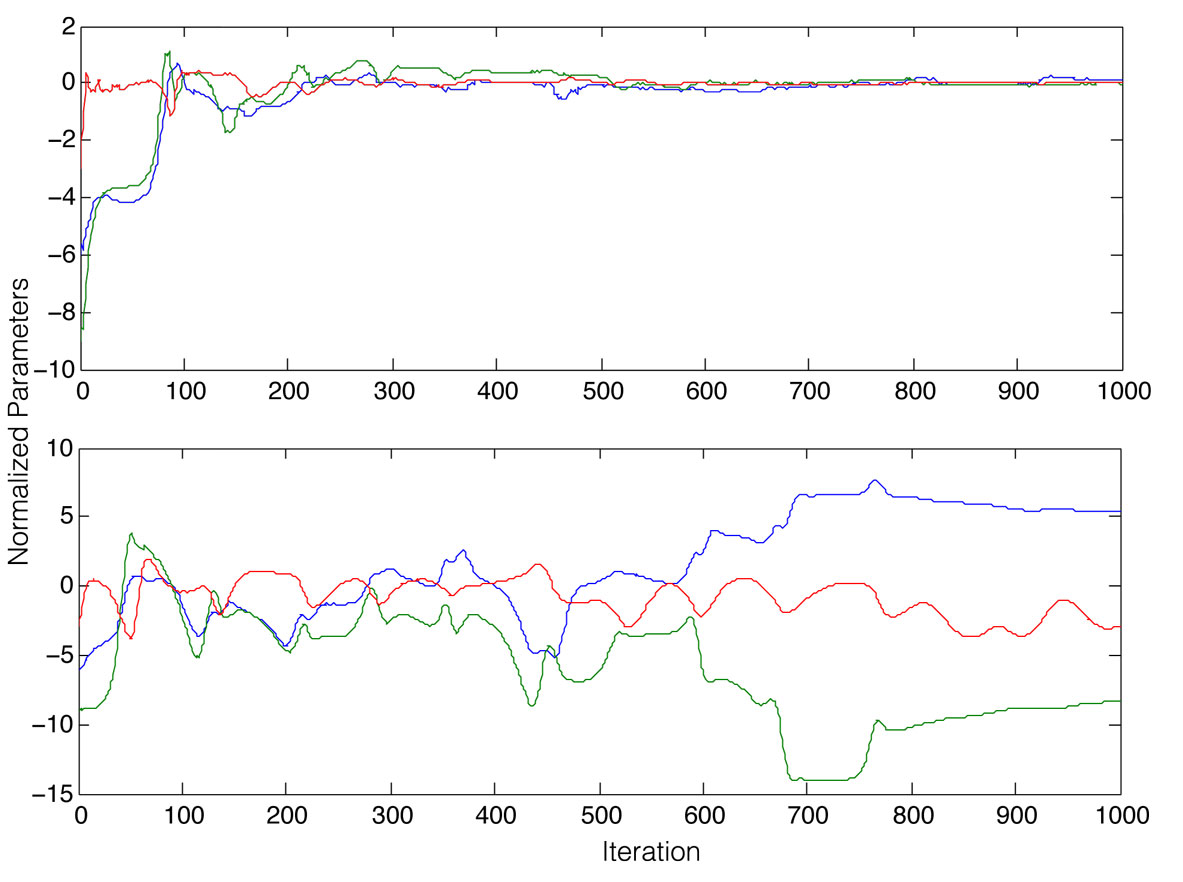}}\\
     \caption{Model parameter convergence}
     \label{fig:params}
\end{figure}

It is evident from the above results that the time to convergence of both the parameter and state depend on the initial conditions.  As such, we conduct a sub-experiment to test model parameter convergence in the Lorenz system.  Specifically, we use DS1 with known noise, and test different initial conditions on the parameters.  We add multiples of the vector $[\sigma_0,r_0,b_0]$ to the true (known) parameters $[10,28,8/3]$ as our initial condition.  Here $\sigma_0=\pm 2$, $r_0=\pm 3$, $b_0=\pm 1$, each with probability $1/2$.  Here we use five different levels of initial conditions corresponding to multiplying the above vector by one through five.  By plotting the average MSE over the three parameters we obtain Figure \ref{fig:paramconv}.  From this we see that for initial conditions close to the true value, convergence is obtained nearly monotonically.  However, an interesting trend is seen for less accurate initial conditions.  Specifically, initially the convergence looks good, then around time 200 the MSE begins to climb.  However, this is not a consistent trend in all of the repetitions.  Rather, this phenomenon is created by a small number of repetitions resulting in parameters which don't converge.  An example of converging (and not converging) parameters is shown in Figure \ref{fig:paramcomp}.  Because of the chaotic nature of the Lorenz system, the parameters often end up on the wrong attractor when poorly initialized, and as a result the parameter estimates do not converge to the true value.  In fact, while most repetitions resulted in convergence of the model parameters, several resulted in similar results as Figure \ref{fig:paramcomp}, with the two variables (blue and green, specify here) staying constant, and the red variable (specify) oscillating in a sinusoidal pattern, indicating that the filter has fallen on a different attractor with similar location as the true solution.

\section{Conclusion}

In this paper, we have explored the effects of misspecification on forecasting methods, where misspecification is defined to include all factors leading to incomplete knowledge of the system at hand, either through lack of knowledge of the system structure, or lack of historical data to learn the system.  We have explored the effects of misspecification on both data and knowledge-based forecasting methods, demonstrating the importance of initialization and adequate levels of historical data.  We have observed that in knowledge-based models where parameters must be learned, care must be taken to ensure sufficient training data is used to ensure accurate parameter estimation.  As the knowledge-based forecaster takes on more aspects of misspecification, the data-based forecaster will outperform it given adequate data.

Through an extensive simulation study, we have demonstrated these notions empirically.  Using the Lorenz-3 dynamic system as an example, we have explored filtering -- a knowledge-based forecasting method, and SVM forecasters -- a data-based forecasting method.  When the filter is provided with complete and accurate information, it dominates the data-based method.  However, when the parameters must be learned and noise is misspecified, the SVM forecaster outperforms the filter.  When stochastic noise is introduced, training data has little positive effect on the forecasters, as there is so little information on the system parameters in the data itself.  If parameters are initialized correctly, convergence and therefore accurate forecasting is likely.  However, when initialized poorly, we have demonstrated that in such chaotic systems as Lorenz-3, the parameters might converge to another mode in the posterior surface of parameters, and hence prediction (particularly long range) will be severely impacted.

\section{Appendix}
\subsection{Appendix 1: Experimental Details of SVM Forecaster}

Let us now describe our experimental setup for the SVM forecaster.  Using an LS-SVM with Gaussian kernel we need to determine both the regularization parameter $\lambda$ and the kernel width $\sigma$.  
To this end, we first scaled the input dimensions, i.e., the past observations,  
of the training set to the interval $[-1,1]$ and saved the corresponding
affine linear transformation so that it can later be applied to the test set.
We further split the training set into four equal parts to perform 4-fold cross validation, and   
considered a grid of candidate values for both $\lambda$ and $\sigma$, where for $\lambda$ we picked 
a grid with endpoints $\lambda_{\mathrm {min}} := 10 (3n/4)^{-2}$ and $\lambda_{\mathrm{max}} := 1.0$, and for $\sigma$ we selected a grid with endpoints $\sigma_{\mathrm {min}} := 0.1$ and $\sigma_{\mathrm{max}} := 2.0 (3n/4)^{1/(3M)}$,
where $n$ is the number of samples in the training set and $M$ is the length of the considered 
history. Here we note that that the behaviour of these endpoints in $n$ and $M$ are suggested by 
theoretical results on support vector machines in the i.i.d.~case, see \cite[Chapter 7.4]{steinwart2008support}.
Though strictly speaking these results do not apply to our datasets, we found that they were actually a good choice 
in all cases. In addition, note that the factor $3/4$ was used to address the fact that in the 4-fold cross validation procedure 
the actual number of samples used for training was approximately $3/4 n$. 
In both cases, 
the grid contained 10 geometrically distributed points.
We then  performed 4-fold cross validation to determine 
the parameter pair $(\lambda^*,\sigma^*)$ from the grids that had the best cross-validation performance.
In other words, for each parameter pair we used three fourths of  the training set for computing the LS-SVM and one fourth
for computing the validation error and repeated this four times with changing validation set. The pair $(\lambda^*,\sigma^*)$ with the smallest average 
validation error was then picked.

This cross-validation procedure  was repeated for all  embedding dimensions $M=1,\dots,50$,
where for $M>5$ we stopped considering higher embedding dimensions if the best cross-validation performance of the current 
embedding dimension was $1.02$ times larger than the best cross-validation performance of all previously considered embedding dimensions.
We then picked the embedding dimensions $M^*$ that had the best cross-validation performance over the  ranges of 
allowed embedding dimensions. For example, if we were allowed to use up to 5 embedding dimensions, we picked the $M^*\in \{1,2,3,4,5\}$
with the best cross validation performance. 
Once we had determined the embedding dimensions $M^*$ and the hyper-parameter pair 
 $(\lambda^*,\sigma^*)$ we then computed the LS-SVM on the entire training set for  $(4/3\lambda^*,\sigma^*)$, where we note that this sort retraining 
 is the standard procedure in machine learning and we had made very good experience with this particualr rescaling of the parameters
 in many previous experiments with SVMs on other datasets.
 For this LS-SVM we then computed the test error with the help of the 
 hold out test set whose input dimensions where scaled by the saved affine linear transformation we found on the training set.

\subsection{Appendix 2: Experimental Details of Filtering Methods}

In each of the Gaussian filters for DS1 through DS6, standard deviation 0.8 is used, and the UKF is used to fit the model.  We also employ the particle filter using Laplace noise $\epsilon_t$ and setting $P(z_0) \sim N(y_1, \Sigma_0)$, where $\Sigma_0$ is chosen to be a diagonal matrix with all elements equal 2.

The choice of proposal distribution has a significant effect on the rate of degeneracy.  The standard (and simplest) choice is the prior distribution $q(z_{t}|z_{0:t-1},x_{1:t}) = P(z_t|z_{t-1})$ since the weights simplify to a calculation of the likelihood.  However, if the likelihood is not near the prior, this choice will lead to large variance in the importance weights, and hence we would like to employ a proposal distribution which uses the data to provide a better estimate of the posterior distribution.  One such possibility is to use a Gaussian approximation of the posterior as the proposal distribution.  We employ the UKF as previously developed to create a proposal distribution for each particle as in \cite{van-der-merwe2001unscented}.  Thus we run the UKF algorithm, but in addition carry a set of particles which are reweighted and resampled based on the UKF foundation.  The situation is complicated in the particle filter situation, due to the well-known problem of degenerate weights (\cite{doucet2001sequential}).  As a solution to this problem, we employ a Gaussian approximation using the UKF for the parameter filter in that case.  In addition, to encourage parameter convergence we employ a tempering effect on $\nu_t$ to shrink it towards zero as a function of time (\cite{liu2001combined}).

Once the end of the observed data $\left\{y_t\right\}_{t\geq1}$ is reached, we end up with a current estimate of state (and perhaps model parameters).  With forecasting in mind, we then use this estimate of state and model parameters $\left( z_T, \theta_T \right)$, combined with the state evolution equation $f(x|\theta)$ to propagate the system foreword. Specifically, the estimate at one time step past the final observation $x_T$ is $z_{T+1}$ = $f(z_T|\theta_T)$, at two time steps past is $z_{T+2} = f(z_{T+1}|\theta_0)$, and so forth.

\bibliographystyle{plain}
\bibliography{lukebornn}

\begin{thebibliography}{10}

\bibitem{aronszajn1950theory}
N.~Aronszajn.
\newblock Theory of reproducing kernels.
\newblock {\em Trans. Amer. Math. Soc.}, 68:337--404, 1950.

\bibitem{black1973pricing}
F.~Black and M.~Scholes.
\newblock {The pricing of options and corporate liabilities}.
\newblock {\em Journal of political economy}, 81(3), 1973.

\bibitem{Cucker2002mathematical}
F.~Cucker and S.~Smale.
\newblock On the mathematical foundations of learning.
\newblock {\em Bull. Amer. Math. Soc.}, 39:1--49, 2002.

\bibitem{Douc2005comparison}
R.~Douc, O.~Capp{\'e}, E.~Polytech, and F.~Palaiseau.
\newblock {Comparison of resampling schemes for particle filtering}.
\newblock In {\em Image and Signal Processing and Analysis, 2005. ISPA 2005.
  Proceedings of the 4th International Symposium on}, pages 64--69, 2005.

\bibitem{doucet2001sequential}
A.~Doucet, N.~{de Freitas}, and N.~Gordon.
\newblock {\em Sequential {M}onte {C}arlo Methods in Practice}.
\newblock Springer, New York, 2001.

\bibitem{julier1997extension}
S.J. Julier and J.K. Uhlmann.
\newblock {A new extension of the Kalman filter to nonlinear systems}.
\newblock In {\em Int. Symp. Aerospace/Defense Sensing, Simul. and Controls},
  volume~3, page~26. Citeseer, 1997.

\bibitem{keerthi2003algorithm}
S.~S. Keerthi and S.~K. Shevade.
\newblock {SMO} algorithm for least squares {SVM} formulations.
\newblock {\em Neural Computation}, 15:487--507, 2003.

\bibitem{liu2001combined}
J.~Liu and M.~West.
\newblock {Combined parameter and state estimation in simulation-based
  filtering}.
\newblock {\em Sequential Monte Carlo methods in practice}, pages 197--223,
  2001.

\bibitem{lorenz1963deterministic}
E.N. Lorenz.
\newblock {Deterministic Nonperiodic Flow.}
\newblock {\em Journal of Atmospheric Sciences}, 20:130--148, 1963.

\bibitem{mcquarrie1997physical}
D.A. McQuarrie and J.D. Simon.
\newblock {\em {Physical chemistry: a molecular approach}}.
\newblock Univ Science Books, 1997.

\bibitem{poggio1990theory}
T.~Poggio and F.~Girosi.
\newblock A theory of networks for approximation and learning.
\newblock {\em Proc. IEEE}, 78:1481--1497, 1990.

\bibitem{sohn2004review}
H.~Sohn, C.R. Farrar, FM~Hemez, DD~Shunk, DW~Stinemates, BR~Nadler, et~al.
\newblock {A review of structural health monitoring literature: 1996--2001}.
\newblock {\em Los Alamos National Laboratory, Los Alamos, NM}, 2004.

\bibitem{steinwart2008support}
I.~Steinwart and A.~Christmann.
\newblock {\em Support Vector Machines}.
\newblock Springer, New York, 2008.

\bibitem{Suykens2002least}
J.~A.~K. Suykens, T.~Van Gestel, J.~{De Brabanter}, B.~{De Moor}, and
  J.~Vandewalle.
\newblock {\em Least Squares Support Vector Machines}.
\newblock World Scientific, 2002.

\bibitem{van-der-merwe2001unscented}
R.~Van~der Merwe, A.~Doucet, N.~De~Freitas, and E.~Wan.
\newblock {The unscented particle filter}.
\newblock {\em Advances in Neural Information Processing Systems}, pages
  584--590, 2001.

\bibitem{Wahba1990spline}
G.~Wahba.
\newblock {\em Spline Models for Observational Data}.
\newblock Series in Applied Mathematics 59, SIAM, Philadelphia, 1990.

\bibitem{wan2000unscented}
E.A. Wan and R.~Van Der~Merwe.
\newblock {The unscented Kalman filter for nonlinear estimation}.
\newblock In {\em The IEEE 2000 Adaptive Systems for Signal Processing,
  Communications, and Control Symposium 2000. AS-SPCC}, pages 153--158, 2000.

\bibitem{wan2000dual}
E.A. Wan, R.~Van Der~Merwe, and A.T. Nelson.
\newblock {Dual estimation and the unscented transformation}.
\newblock {\em Advances in Neural Information Processing Systems}, 12:666--672,
  2000.

\end{thebibliography}

\end{document}